\documentclass[aps,pre,twocolumn,amssymb]{revtex4}
\pdfoutput=1

\usepackage{graphicx}
\usepackage{dcolumn}
\usepackage{bm}

\def\be{\begin{equation}}
\def\en{\end{equation}}
\def\bea{\begin{eqnarray}}
\def\ena{\end{eqnarray}}
\def\bec{\begin{equation}\begin{array}{rcl}}

\def\ls{\lesssim}
\def\ve{\varepsilon}

\newcommand{\av}[1]{\langle{#1}\rangle}

\newcommand{\bi}[1]{\mbox{\boldmath$#1$}}

\begin{document}
\title{ Structure Formation due to Antagonistic Salts }
\author{Akira Onuki$^{a}$\footnote{Corresponding author.\\ 
Email address:  onuki@scphys.kyoto-u.ac.jp (A. Onuki)},
  Shunsuke Yabunaka$^b$, Takeaki Araki$^a$, Ryuichi Okamoto$^c$ }
\address{$^a$ Department of Physics, Kyoto University, Kyoto 606-8502,
Japan\\
$^b$ Yukawa Institute for Theoretical Physics, Kyoto University, Kyoto 606-8502, Japan\\
$^c$ Department of Chemistry, Tokyo Metropolitan University, Hachioji, 
Tokyo 192-0397, Japan
}


\date{\today}

\begin{abstract}
Antagonistic salts are  composed of    hydrophilic and  hydrophobic 
ions.  In a  mixture solvent (water-oil) 
 such ion pairs   are  preferentially attracted  to water or oil, 
giving rise to a  coupling between 
the  charge density and the composition. 
First, they form a large electric double layer 
at a water-oil  interface,  
reducing the surface tension and producing  mesophases. 
Here,  the cations and anions are loosely bound  
by  the Coulomb attraction   across the interface 
on the scale of the Debye screening length. 
Second,   on solid surfaces, 
hydrophilic (hydrophobic) ions 
are  trapped  in a water-rich (oil-rich) 
 adsorption layer, while those  of the other 
species are  expelled  from the layer. This  yields  
a solvation mechanism of local charge separation near a solid.   
In particular, near the solvent criticality,  
disturbances around solid surfaces can  
 become oscillatory in space. 
In mesophases, we calculate    periodic structures, 
which resemble  those in experiments. 
\end{abstract}


\maketitle


\section{Introduction}
\setcounter{equation}{0}

In soft materials, 
much attention has been paid to the  Coulomb interaction 
among  charged objects.   
 However, not enough attention 
 has yet been paid to    the solvation (or hydration) 
 interaction between ions  
and polar solvent molecules  \cite{Is}, particularly when 
ions induce  some phase ordering \cite{Bu}. 
In liquid water,  small metallic   ions  are surrounded by 
several  water  molecules to form  a hydration  shell 
  due to the ion-dipole  interaction. 
 As a result, such  ions are  strongly hydrophilic. 
On the other hand,  hydrophobic ions have been used  
 in electrochemistry \cite{Bu}. An example is  tetraphenylborate BPh$_{4}^-$   
with a diameter about 0.9 nm  
consisting     of  four phenyl rings bonded to a negatively ionized 
boron (see Fig.1).  Because of its large size, it largely deforms  
the surrounding hydrogen bonding   acquiring  strong hydrophobicity 
\cite{Chandler}.

In this review, we explain some  unique effects 
emerging  when  both hydrophilic and hydrophobic ions 
are present in a mixture solvent 
composed of  water   and  a less polar 
component (called oil). In this situation, such cations and anions 
 behave {\it antagonistically} in the presence of composition 
heterogeneity  and   even induce mesophases with charge density waves. 
As in Fig.1,  they tend to 
segregate  around a  water-oil interface 
due to the selective solvation,  but they can only  undergo 
microphase separation  on the scale of the Debye 
screening  length due to the Coulomb attraction \cite{OnukiPRE}. 
Indeed, in  a x-ray   reflectivity experiment,  
Luo {\it et al.} \cite{Luo} observed such ion distributions   
around a water-nitrobenzene interface.
The  resultant electric double layer reduces 
the surface tension $\gamma$ \cite{OnukiJCP,Araki,Nara}. 
Such a decrease in $\gamma$ 
was recently observed\cite{Bonn}.
Thus, with an antagonistic salt 
in the vicinity of the solvent criticality, 
 we have $\gamma<0$  to find   mesophases.

 Sadakae {\it et al.} \cite{Sadakane,Sada1} 
detected a peak in the small-angle neutron scattering 
amplitude at an intermediate wave number  
  adding   a small amount of 
NaBPh$_{4}$   in  a  near-critical 
D$_2$O  and 3-methylpyridine (3MP) mixture. 
Such a peak indicates  mesophase formation.  
In the phase diagram of D$_2$O-3MP,  
 the closed-loop  two-phase region shrinks 
  with   NaBPh$_{4}$, while it expands with  
 a hydrophilic salt such NaCl\cite{Sada1}. 
 Moreover, they  observed  multi-lamellar (onion) 
structures   at  small volume fractions of 3MP 
  far from the criticality \cite{SadakanePRL,Sada2}. 
Afterwards, Leys {\it et al.} \cite{Leys}   
 examined  the dynamics in D$_2$O-3MP with  NaBPh$_{4}$ 
using dynamic light scattering and small-angle neutron scattering.

We also point out    that 
hydrophilic (hydrophobic) ions can be 
selectively adsorbed into a 
water-rich (oil-rich) adsorption layer 
on a solid wall. This effect  is intensified 
for  antagonistic salts, since the 
disfavored ions are expelled  from the  layer due to 
the solvation interaction. 
Its  thickness  is microscopic far from the 
 criticality but is widened in its vicinity.
The adsorbed ion  amount   can   exceed    
the bare surface  charge of the wall  in opposite sign. 
This is  a   chemical mechanism of charge inversion 
\cite{Lyk,Faraudo1,Faraudo}, 
 originating from the selective solvation.

Another interesting effect with  antagonistic salts 
is that the disturbances  
around solid surfaces become oscillatory in space 
near   the transition to mesophases
\cite{Oka,Ciach}. This gives rise to oscillatory 
dependence of the  force  between two colloidal   
particles or two parallel 
walls  on their  separation distance $d$. Furthermore, 
dynamics  such as  
phase ordering and rheology 
is of great interest. 
Thus, more experiments are very informative.

In Sec.2,  we will present   the background of physics and 
chemistry in aqueous mixtures with antagonistic ion pairs. 
In Sec.3, we will treat  salt-induced mesophases.

\section{ Theoretical background}
\noindent{\it{2.1. Solvation chemical potential } }\\
For   each  solute  particle, the solvation part of the 
chemical potential is written as 
$\mu_{\rm s}^i(\phi)$. 
In  mixture solvents,  it depends on the 
 water composition $\phi$. 
Let us suppose  two species of monovalent ions ($Z_1=1$, $Z_2= -1$). 
At sufficiently low ion densities, 
the  total ion chemical potential $\mu_i$ in a  mixture solvent  
is expressed as 
\be 
\mu_i= k_BT \ln (n_i\lambda_i^3) + Z_i e
\Phi+ \mu_{\rm s}^i (\phi),
\en 
where $\lambda_i$ is the thermal de Broglie length   
and $\Phi$  is the  electric potential. 
In equilibrium  $\mu_i$ is homogeneous. 
The  $\phi$ dependence of  $\mu_{\rm s}^i (\phi)$ 
   has not been well investigated.

We consider  a liquid-liquid  interface  between  
a  water-rich phase $\alpha$ and an oil-rich  
phase $\beta$ with bulk compositions $\phi_\alpha$ 
and $\phi_\beta$ and with bulk ion densities $n_{i\alpha}$  and $n_{i\beta}$. 
For each ion species $i$ we introduce   the difference, 
\be 
\Delta\mu_{\rm s}^{i}
= \mu_{\rm s}^{i}(\phi_\alpha)-\mu_{\rm s}^{i}(\phi_\beta),  
\en 
which is the Gibbs energy of transfer in electrochemistry 
(usually measured in units of kJ$/$mol)\cite{Marcus,Hung,Koryta,Osakai}. 
It is negative (positive) for hydrophilic (hydrophobic) ions. 
The electric potential    $\Phi$ tends to 
 constants  $\Phi_\alpha$ in $\alpha$ 
and $\Phi_\beta$ in $\beta$, yielding   a  Galvani potential difference, 
\be 
 \Delta\Phi=\Phi_\alpha-\Phi_\beta=   ({\Delta\mu_{\rm s}^2 
-\Delta\mu_{\rm s}^{1}})/2e,
\en 
From the charge neutrality in the bulk, we have 
\be 
{n_{i\beta}}/{n_{i\alpha}}
= \exp[-({\Delta\mu_{\rm s}^{1}} +
 \Delta\mu_{\rm s}^{2} )/{2k_BT} ].
\en 

 For  water-nitrobenzene (NB) 
at $T=300$ K in strong segregation \cite{Bu,Hung,Koryta,Osakai},
  $\Delta\mu_{\rm s}^i/k_{B}T$ was  
$-13.6$  for Na$^+$,   $-27.1$  for Ca$^{2+}$, and 
$-11.3$ for Br$^-$, while it was $14.4$ for BPh$_4^-$. In 
 the NB-rich phase in this case, the water content is 
$\phi\sim 0.003 $  and  well-defined 
 hydration shells are already formed around metallic ions\cite{Osakai}. 
 For water-alcohol \cite{Marcus}, 
$\Delta\mu_{\rm s}^i$ was observed to be relatively small 
in weak segregation. 
 For hydrophobic particles, 
$\mu_{\rm s}^i(\phi)$ depends on how they deform 
the surrounding hydrogen bonding, so 
it increases sharply with increasing their size\cite{Chandler}.

\begin{figure}
\begin{center}
\includegraphics[width=0.9\linewidth]{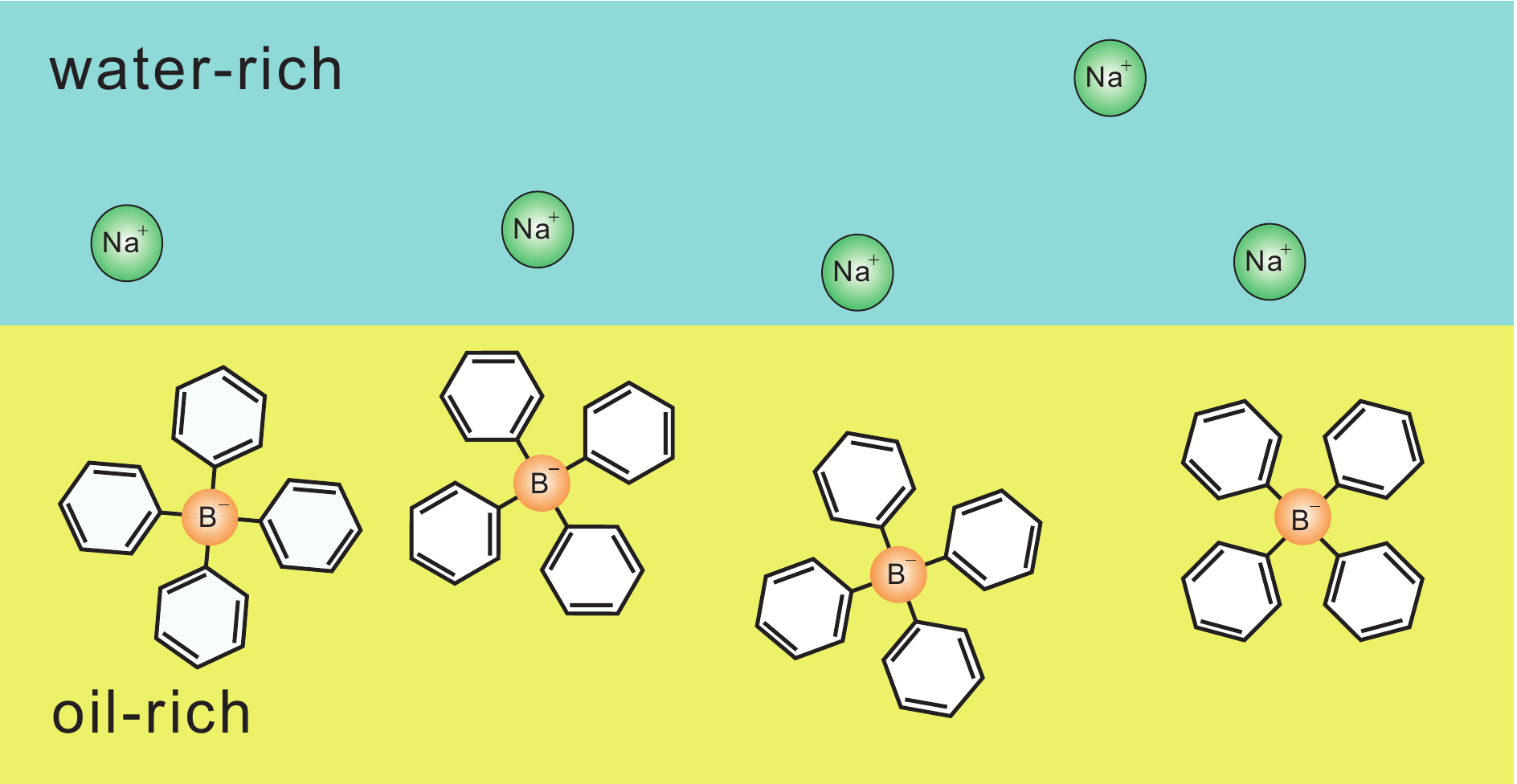}
\caption{Hydrophilic  Na$^+$  in water-rich region and 
hydrophobic BPh$_4^-$  in oil-rich region  around  an interface.  } 
\end{center}
\end{figure}

\begin{figure}[t]
\begin{center}
\includegraphics[scale=0.38]{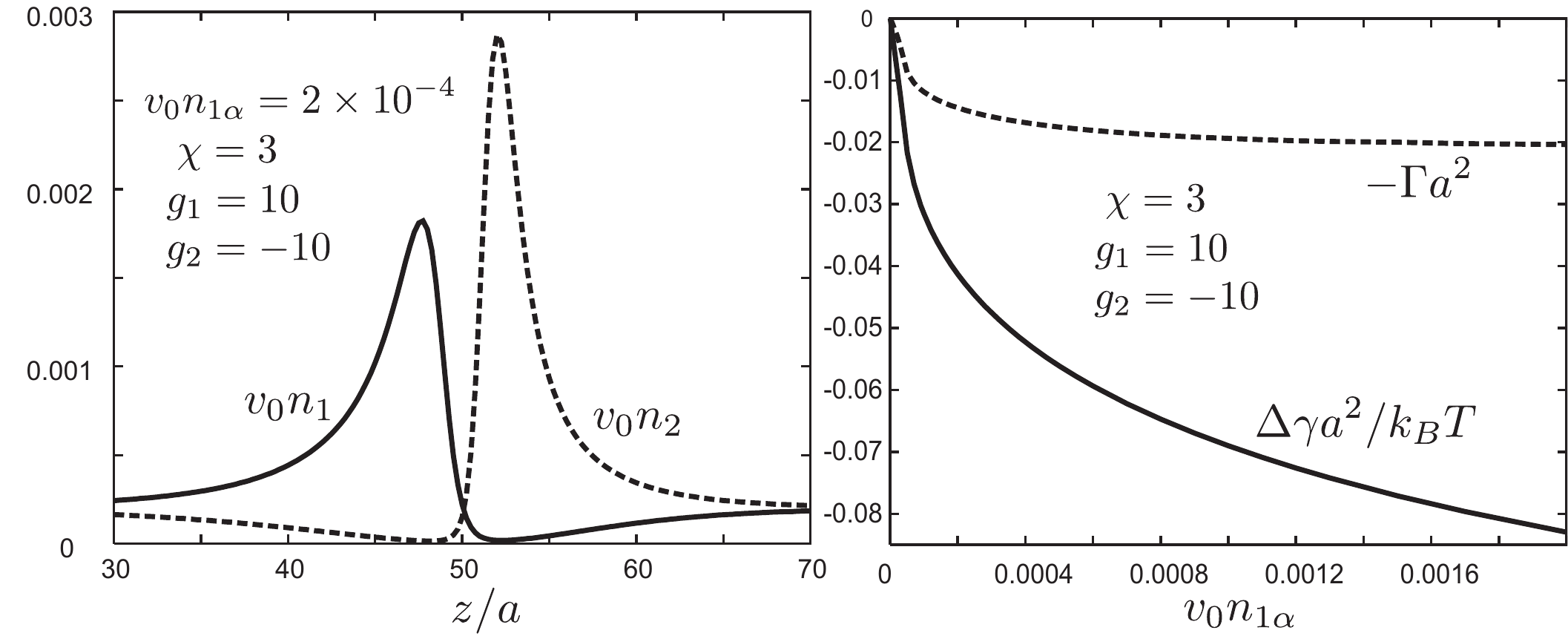}
\caption{ 
Left: Interfacial ion densities $n_1(z)$ and $n_2(z)$ 
 multiplied by $v_0= a^3$  vs $z/a$ 
 for  antagonistic pair, where $v_0 n_{1\alpha} =
4 \times 10^{-4}$.  Right: Surface tension decrease 
$\Delta\gamma=\gamma-\gamma_0$ divided by $k_BT/a^2$ 
 and surface ion adsorption $\Gamma$ multiplied by $-a^2$ 
 vs $v_0 n_{1\alpha}$. Here, $\chi = 3$  and $g_1 = -g_2 = 10$.
}\end{center}
\end{figure}

\vspace{2mm}
\noindent{\it{2.2.Ginzburg-Landau theory}}\\
In our previous papers \cite{OnukiPRE,OnukiJCP,Araki,Nara}, 
we used  the Ginzburg-Landau theory \cite{Onukibook} to examine 
  the interface profiles and the phase behavior,  
 where $\phi$ and $n_i$ are 
coarse-grained variables. The free energy density is  
the sum of the electrostatic energy 
$\ve(\phi) |{\bi E}|^2/8\pi$ and $f_{\rm tot}$ 
given by 
\bea 
\frac{f_{\rm tot}}{k_BT}
&=& \frac{1}{v_0} [\phi \ln\phi + (1-\phi)\ln (1-\phi) 
+ \chi \phi (1-\phi)]\nonumber\\
&&\hspace{-0.5cm}+ \frac{1}{2}C|\nabla\phi|^2+ 
\sum_i n_i [\ln (n_i\lambda_i^3) -1-  g_i \phi], 
\ena 
where $v_0=a^3$ is the molecular volume 
with $a \sim 3~{\rm \AA}$, $\chi(T)$ is the 
interaction parameter  depending on $T$, and 
we set $C=a^{-1}$. Though very crude, we use the linear form 
$
\mu_{\rm s}^i(\phi) = A_i -k_BT g_i \phi, 
$ 
where  $A_i$ is   a constant. 
Then,   $\Delta\mu_{\rm s}^{i}= - k_BT g_i (\phi_\alpha-\phi_\beta)$. 
In our theory, $g_i$ represents the solvation  strength.
 The electric field  ${\bi E}=-\nabla\Phi$ satisfies the  
 Poisson equation $
\nabla\cdot\ve{\bi E}/  4\pi=\rho= e(n_1-n_2)$, where the 
  dielectric constant  depends on $\phi$ as 
 $\ve(\phi)=\ve_0+\ve_1\phi$. In this work,  we set 
$\ve_1=\ve_0$ and $e^2/\ve_0 k_BT =3a$. 

For large $|g_i|\gg 1$, the phase behavior from our model  
is  complicated even for small $n_i$.  
See the mean-field phase diagram in our previous paper\cite{Nara}. 
 Let a  homogeneous solution have  mean 
water concentration $\phi$ and mean 
ion densities $n_1=n_2=n_e$. A  linear instability occurs  for  
\be 
\tau < (g_1+g_2)^2{n_e}/2 +8\pi \ell_B  C(\gamma_{\rm p}-1)^2n_e .  
\en 
Here,  $\tau= [1/\phi(1-\phi)-2\chi]/v_0$ is the second derivative 
of the first term in Eq.(5) with respect to $\phi$, 
  $\ell_B=e^2/\ve k_BT$ is  the Bjerrum length, and  the parameter 
\be 
\gamma_{\rm p}= (16 \pi C \ell_{B})^{-1/2} |g_1-g_2|  
\en 
represents  the strength of antagonicity. The Debye wave number 
is $\kappa
 = (8\pi \ell_B n_e)^{1/2}$. If  Eq.(6) 
holds with  $\gamma_{\rm p}>1$, the fluctuations  with  
wave number $q_{\rm m}= 
\kappa (\gamma_{\rm p}-1)^{1/2}$ 
grow to form  a  periodic structure. 

\begin{figure}[t]
\begin{center}
\includegraphics[scale=0.49]{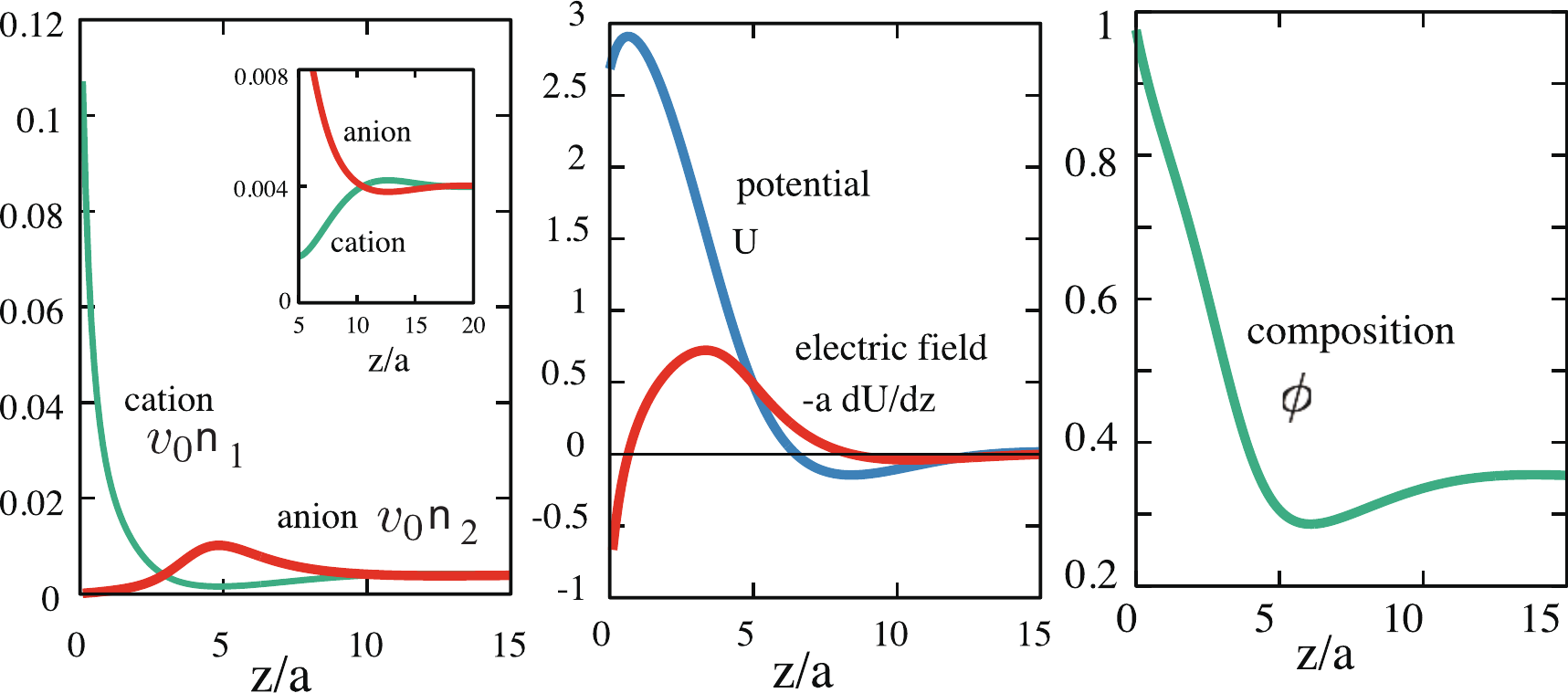}
\caption{ Profiles  near  hydrophilic wall 
with surface charge density $\sigma_0= 
-0.04e a^{-2}$ at $z=0$, where $\chi=2$,  $g_1=-g_2=10$, 
 and  $v_0 n_i\to 4 \times 10^{-3}$ as $z\to\infty$.  
(a)  Ion densities, where cations are adsorbed near the wall  
 in excess of $\sigma_0<0$. They are expanded 
in region $5<z/a<20$ (inset). (b) 
Normalized potential $U=e\Phi/k_BT$ and electric field $- a dU/dz= (ea/k_BT)E$. 
 (c)  Composition $\phi$,  tending  to  $ 0.35$ as $z\to \infty$. }
\end{center}
\end{figure}
\begin{figure}[t]
\begin{center}
\includegraphics[scale=0.42]{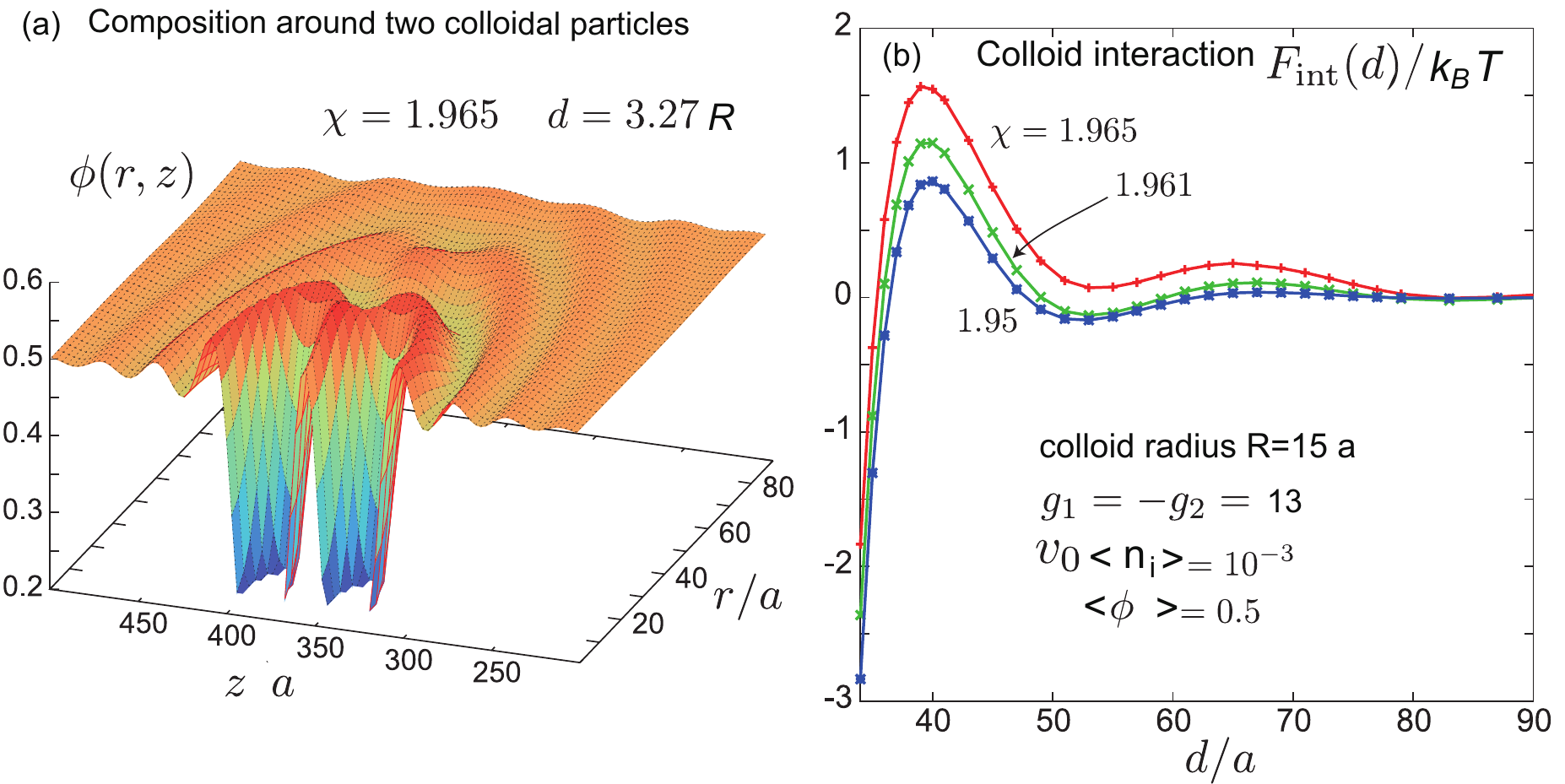}
\caption{ Two hydrophobic colloidal 
particles with radius $R=15a$ 
and  separation  $d=3.27R$ without surface charges 
\cite{Oka}, where $\chi=1.965$,  
 $g_1=-g_2=13$, $v_0\av{n_i}=10^{-3}$,  
and $\av{\phi}=0.5$. 
(a) Composition $\phi(r,z)$ in the $r$-$z$ plane, where 
$r= (x^2+y^2)^{1/2}$.  
(b) Normalized interaction free energy $F_{\rm int}(d)/k_BT$ 
vs  $d/a$ for  $\chi=1.950$, $1.961$, and $1.965$. 
   }
\end{center}
\end{figure}

\begin{figure}[t]
\begin{center}
\includegraphics[scale=0.46]{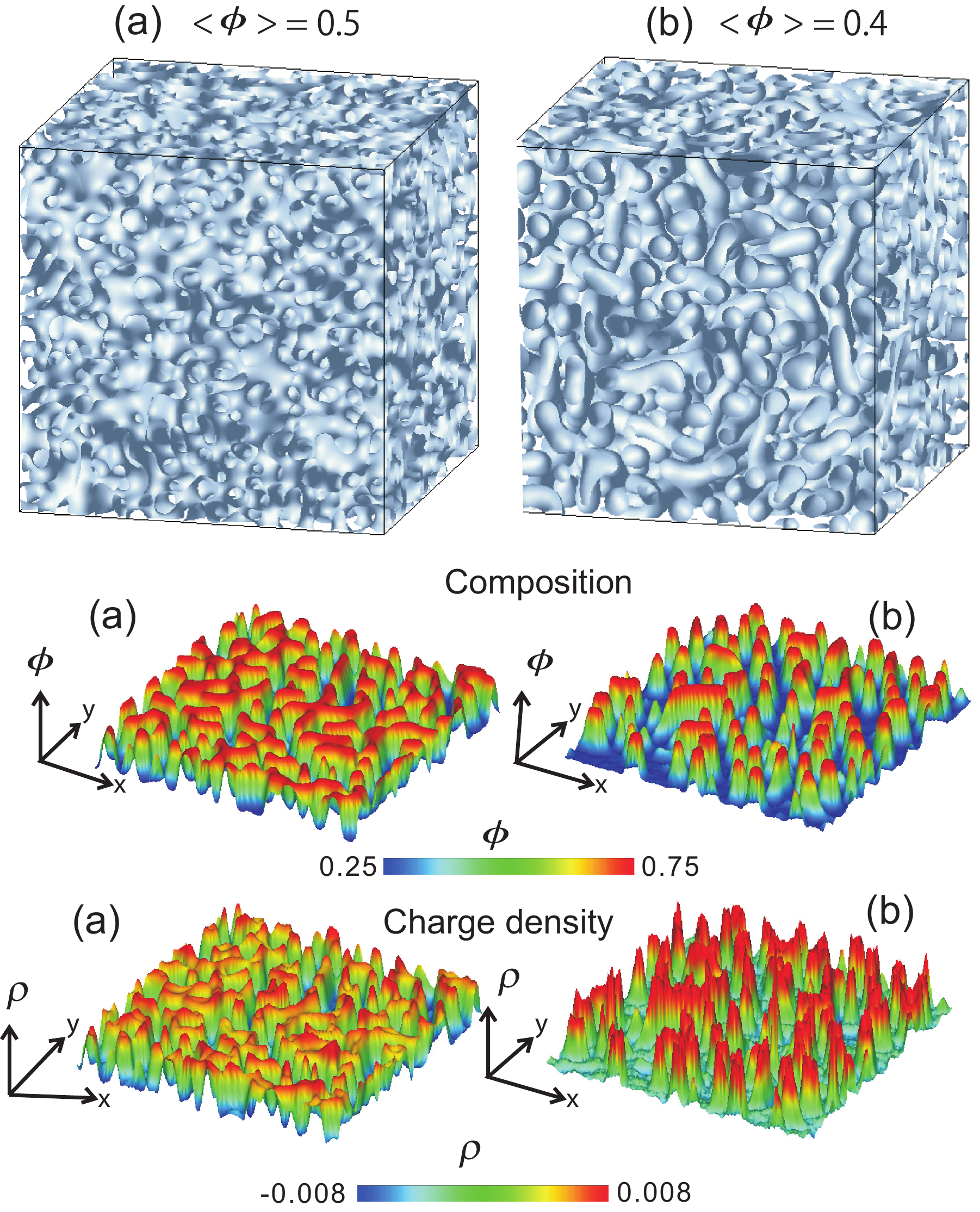}
\caption{Mesophase patterns    near solvent criticality  
induced by   antagonistic salt. They are calculated in a box 
of length $128a$ with    $\chi=2$, $g_1=-g_2=10$, 
 and  $v_0\av{n_i} =7.3\times 10^{-4}$. 
Average composition  $\av{\phi}$ is (a) 0.5 (left) and (b)0.4 (right).  
These patterns are  nearly stationary without thermal noises. 
Top: Surfaces of $\phi({\bi r})=0.5$, 
which  are (a) bicontinuous  and (b) tubelike.
Middle: Cross-sectional profiles  of composition  $\phi(x,y,0)$. 
Bottom: Those  of charge density  $\rho (x,y,0)$.  The 
colors represent $e\Phi/k_BT$ 
according the color bar.}
\end{center}
\end{figure}

\vspace{2mm}
\noindent{\it{2.3.Interface  profiles }}\\
In  Fig.2, we display  typical ion profiles 
near an interface (left) along the $z$ axis, where hydrophilic cations 
 and hydrophobic anions are separated on 
the  scale of the Debye  length $\kappa^{-1}(\sim 10a)$.  In the right panel, 
we show the surface tension 
deviation $\Delta\gamma=\gamma-\gamma_0$ 
and the ion adsorption $\Gamma$ vs  the ion density $n_{1\alpha}$ 
in phase $\alpha$. Here,   $\gamma_0$ is the surface tension without ions 
and $\Gamma$ is  the integral of 
$n(z)-n_\alpha-(\phi(z)-\phi_\alpha)(n_\beta-n_\alpha)/
(\phi_\beta-\phi_\alpha)$ 
with $n=n_1+n_2$. For small  ion densities, $\gamma$  is 
given by \cite{OnukiJCP,OnukiPRE} 
\be 
\gamma\cong  \gamma_0- k_BT \Gamma - \int dz \ve |{\bi E}|^2/8\pi . 
\en   
The third electrostatic term is of order 
 $-\kappa \ve_0 |\Delta\Phi|^2/4\pi$ 
in terms $\Delta\Phi$ in  Eq.(3),   
dominating over the second Gibbs term. 
We may  realize  $\gamma<0$ 
with increasing the salt amount and/or 
approaching  the solvent criticality.

\vspace{2mm}
\noindent{\it{2.4. Selective  ion adsorption on a wall}}\\
In Fig.3, we plot profiles near a hydrophilic wall, 
where we impose   the boundary condition  $d \phi/dz= -0.2/a$ 
at $z=0$. Here,  the cations 
are accumulated in the adsorption layer
($z/a\ls 4$), while the anions are 
expelled from it. For simplicity, we neglect the image interaction. 
The electric field $E$ along the $z$ axis is given by 
\be  
\ve(z) E(z)/ 4\pi = \sigma_0 + \int_0^z dz'\rho(z').
\en  
The  surface charge density is $\sigma_0= -0.04e a^{-2}$. 
As a result, we find $E<0$ for $z/a<0.7$ and  
$E>0$ in an intermediate 
range  $0.7 <z/a< 8.4$. For larger $z$, $E$ is very small and 
is oscillating. 

 Thus, both  adsorption and desorption of ions 
take place simultaneously   on solid  surfaces 
with  antagonistic salts in mixture solvents. 
However, highly hydrophobic ions such 
 AsPh$_{4}^+ $(tetraphenylarsenate)
\cite{Faraudo} 
and BPh$_{4}^-$ \cite{Faraudo1} are    absorbed on hydrophobic 
 colloid surfaces  in  water (without oil) 
 leading to charge inversion.

\vspace{2mm}
\noindent{\it{2.5.Spatially oscillating disturbances }}\\
Let us assume that  $\tau$ is slightly larger than the right hand side 
 with $\gamma_{\rm p}>1$ in Eq.(6)  \cite{Oka,Ciach}. 
In this case,   the deviations of the composition and 
the electric potential are strongly coupled. 
As a result,  oscillatory  disturbances 
appear around  colloidal particles 
\cite{Oka} or between parallel solid walls \cite{Ciach}  in equilibrium.

In Fig.4, we show a wavelike profile  
$\phi(r,z)$ around two colloidal particles 
with radius $R=15a$  at $\chi= 1.965$\cite{Oka}, 
 where  $r= (x^2+y^2)^{1/2}$. 
 The center-to-center distance is 
 $d=3.27R$. Their  surfaces are surrounded by oil layers 
rich in  hydrophobic anions without surface charges. 
They are effectively charged as in Fig.3.   
No van der Waals interaction is assumed. 
 In the right panel, 
 the interaction free energy $F_{\rm int}(d)$ 
of the two particles is plotted as a function of 
 $d$.  Instability to a mesophase occurs for  $\chi>1.9651$, 
so  homogeneity is attained far from the particles.
Here,  $F_{\rm int}(d)$ exhibits a shallow minimum  
at $d\sim 80 a$ and 
 a  maximum of  order $k_BT$ at $d\sim 40 a$. 
The  maximum  increases as $R^2$ 
with increasing the  radius $R$, 
so it  can   prevent  
 contact of the particles for large  $R$.

\begin{figure}[t]
\begin{center}
\includegraphics[scale=0.59]{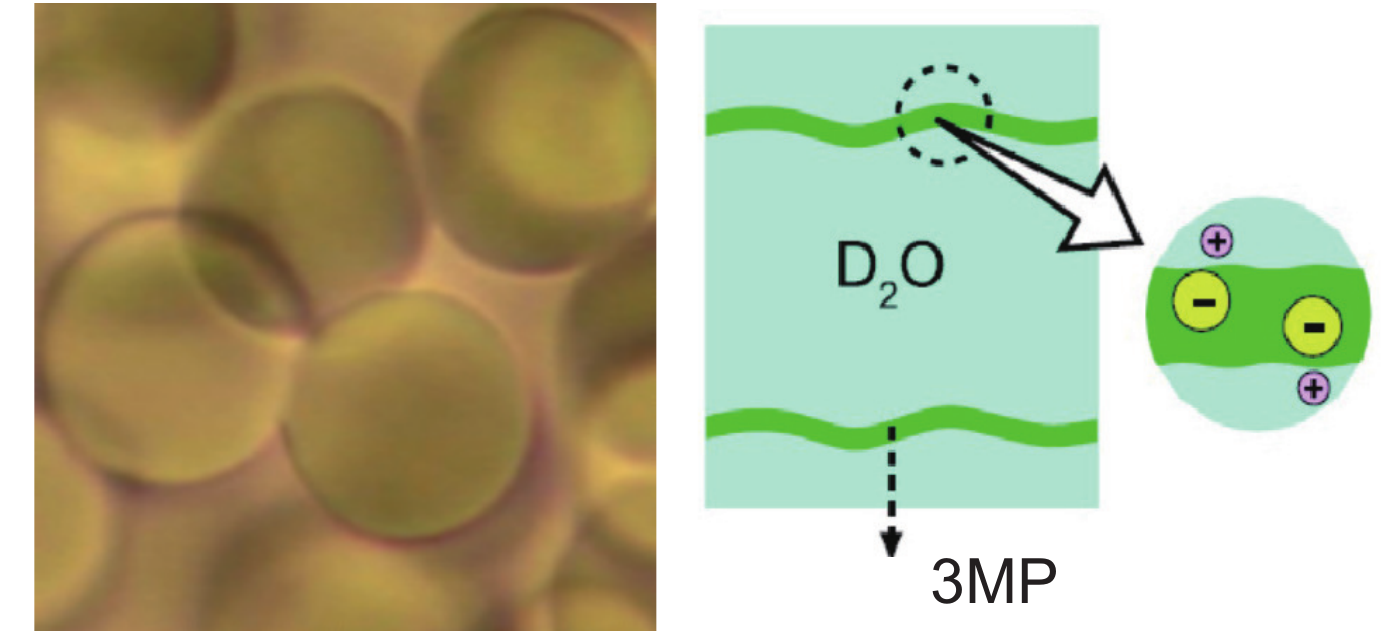}
\caption{ Left: Optical microscopic image 
of onions with radius about $10^4$ nm 
at   $T=313$ K in  D$_2$O-3MP with NaBPh$_{4}$ (0.085 mol$/$L) 
far from criticality,  
where  D$_2$O   volume fraction is 0.91  
(taken from Ref.\cite{SadakanePRL}). 
Right: Illustration of lamellar structure, 
where spacing is $17.5$ nm 
and  membranes consisting of 
3MP and BPh$_{4}^-$  have  thickness about $1.6$ nm.  
 }
\end{center}
\end{figure}

\begin{figure}[t]
\begin{center}
\includegraphics[scale=0.48]{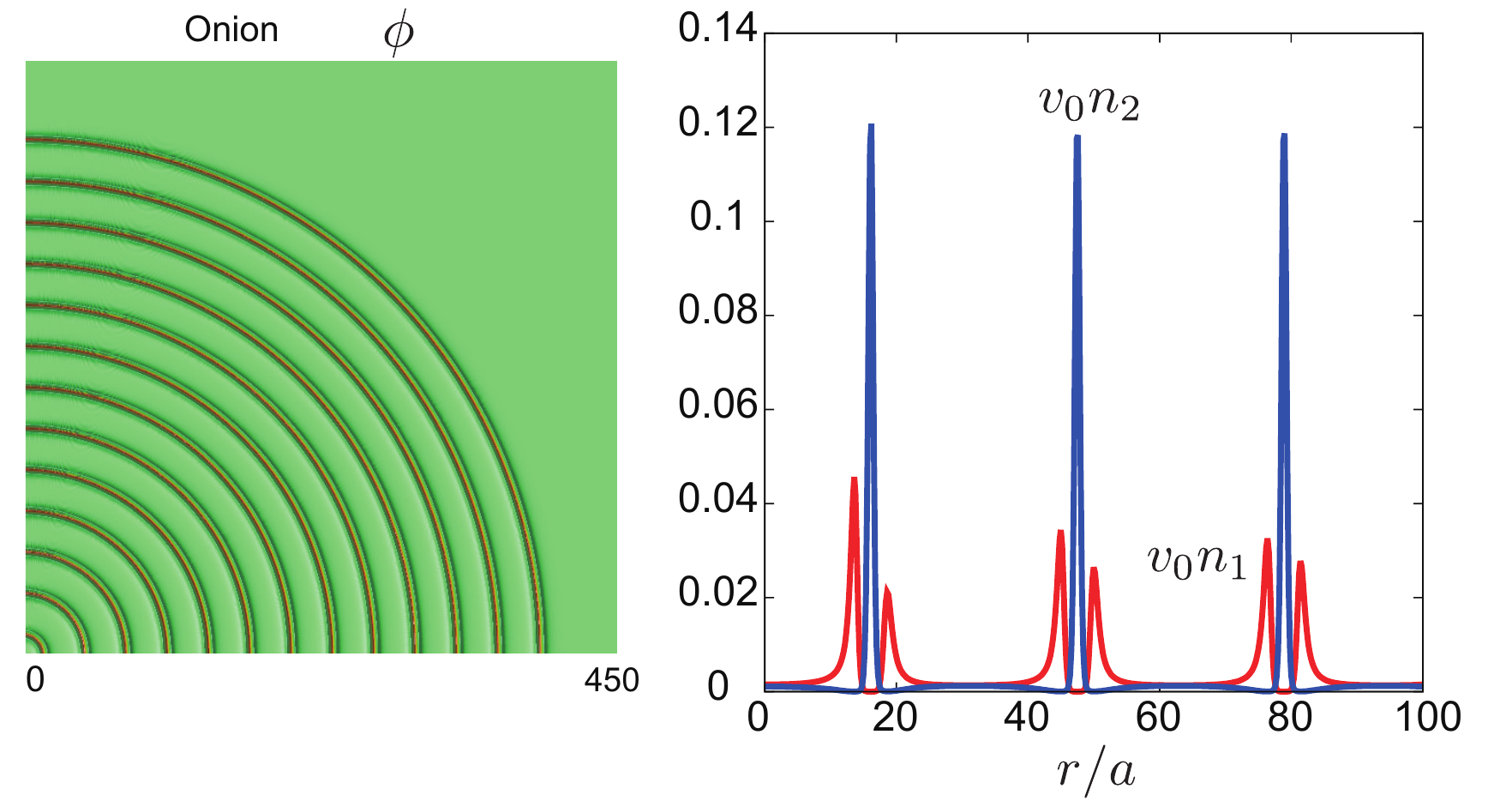}
\caption{ Numerically calculated onion,  
    where $\chi=2.25$, $g_1= - g_2=15$,  $\av{\phi}=0.8$,  
 and  $v_0\av{n_i} =3\times 10^{-3}$. 
Membranes are composed of  oil and hydrophobic anions.
(a) Cross-sectional profile of composition $\phi(r)$ 
($r$ is the distance from the center).
(b) Those of $v_0 n_1(r)$ and $v_0n_2(r)$ 
around 3 membranes. 
 }
\end{center}
\end{figure}

\begin{figure}[t]
\begin{center}
\includegraphics[scale=0.55]{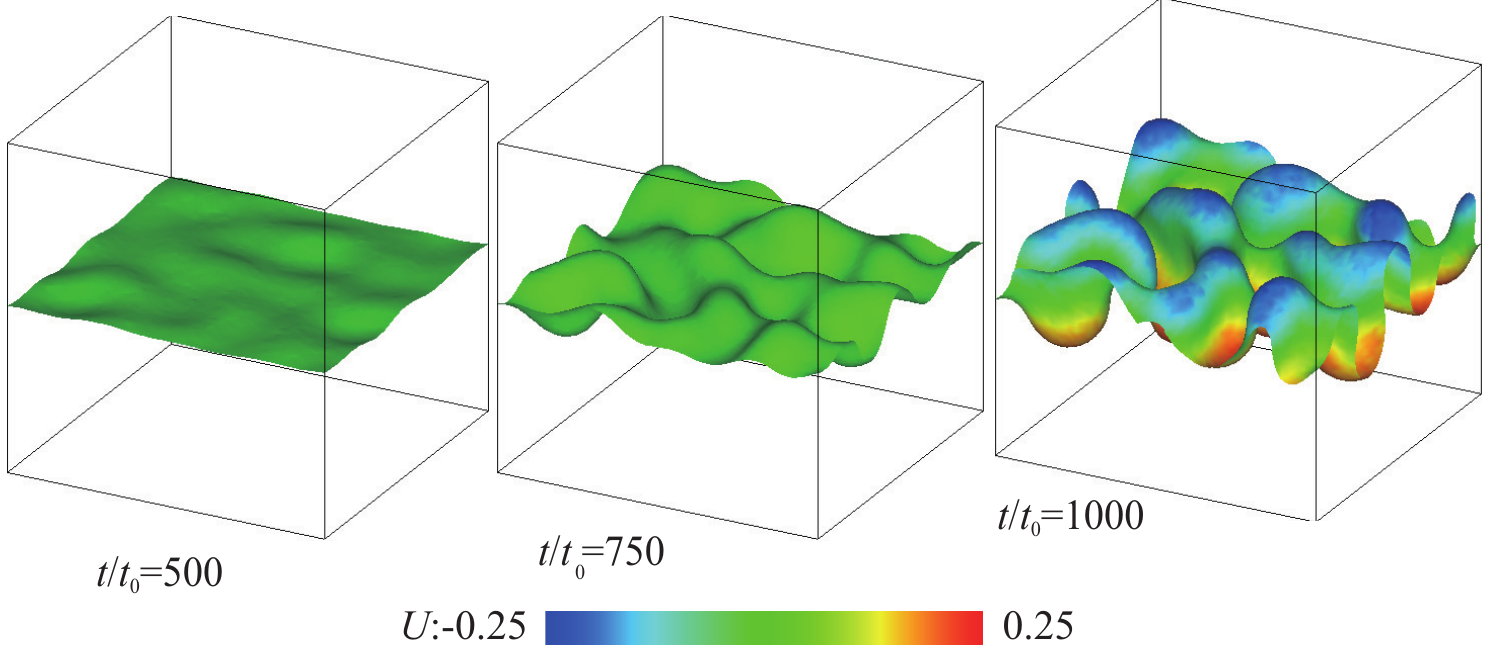}
\caption{Growth of interface deformations induced by 
antagonistic salt \cite{ArakiJ} in a box of length $64a$  
at $t/t_0=500, 750$, and 1000, where 
$t_0= 2\pi  a^3\eta/k_BT$.  The ion diffusion constant is 
$D= a^2/t_0$ for the two species.  
Here,  $\chi=2.1$  and  $g_1=-g_2=12$. 
  The  colors represent $U=e\Phi/k_BT$. }
\end{center}
\end{figure}

\section{Mesophase formation}

\noindent{\it{3.1. Mesophases near criticality}}\\
To calculate   mesophase structures near the 
 criticality\cite{Sadakane,Sada1},  
we changed $\chi$ from a small value 
to $\chi_c= 2$ at 
$t=0$, where   Eq.(6) is satisfied with $\gamma_{\rm p}=1.4$. 
We used model H equations with  viscosity $\eta$ 
\cite{Araki,Nara,Onukibook}. The random 
patterns in Fig.5 are those at 
$t= 1000t_0$ with $t_0= 2\pi a^3 \eta /k_BT$, which  
 are nearly stationary without thermal  noises    
in the  dynamic equations. The domain sizes are on the order of 
the Debye length $\kappa^{-1}$. 
For  (a) $\langle \phi\rangle=0.5$,  
we can see  bicontinuous patterns,  resembling those 
in spinodal decomposition at a critical quench  
\cite{Onukibook}. 
For (b) $\langle \phi\rangle=0.4$, 
tubelike  domains appear, where 
the droplet  volume fraction (of  the region 
$\phi>1/2)$ is  0.25.  Remarkably, 
the droplets are considerably elongated. 
The  degree of elongation 
increases  with  increasing the salt  amount 
(not shown in this paper).  
In the lower panels, 
we give the cross-sectional profiles  of $\phi$ and $\rho=e(n_1-n_2)$. 
We can see that  the  domains with 
$\phi>\langle \phi\rangle$ ($\phi<\langle \phi\rangle$) 
 are positively (negatively) charged. 

With further increasing $\chi$  inside 
the  two phase region, 
a changeover should   occur from random mesophases into 
macroscopic two-phase states.  
In real near-critical systems, these mesophase fluctuations 
 are  thermally  changing   in time and their influence  
 on the  static and dynamic critical behaviors 
remains unclear \cite{Sada1,Leys}.

\vspace{2mm}
\noindent{\it{3.2. Onions}}\\
Second, as illustrated in Fig.6, 
Sadakane {\it et al.} \cite{Sada2,SadakanePRL} 
observed onions of many membranes ($\sim 500)$ 
 composed of 3MP and  BPh$_{4}^-$. 
The  lamellar   spacing is  $17.5$ nm 
and the membrane thickness is  $1.6$ nm. 
These onions are due 
to strong hydrophobicity of BPh$_{4}^-$ 
in a water-rich enviroment.
The  membranes are analogous to those 
of ionic surfactants 
and the adsorbed  BPh$_{4}^-$ density 
on them is of order $0.01~{\rm \AA}^{-2}$.  
.

We have also performed simulation to find  a similar 
onion in Fig.7, where  
 $\chi=2.25$, $g_1= - g_2=15$,  and 
$\av{\phi}=0.8$. 
We have minimized the free energy  
 in the spherically symmetric geometry 
in the region   $0<r<500a$, where $r$ is the distance from 
the center.    The lamellar spacing and thickness are 
$32 a$ and  $6a$, respectively. The adsorption 
of the anions  within each  membrane 
 is $ 0.116 a^{-2}$ per unit area. 
The composition $\phi(r)$ 
 is 0.827 outside  the membranes 
(commonly within and outside  the onion), while 
the average composition within 
the membranes becomes about 0.5.

.  

\noindent{\it{3.3. Interface instability}}\\
In Fig.8,   we also  numerically examine 
the  interface  instability \cite{ArakiJ} 
 with $\chi=2.1$ and $  g_1=-g_2=12$ 
in a cubic cell with length $L=64 a$. 
At $t=0$, an interface was at $z= L/2$ 
and  we   added    
 antagonistic ion pairs 
to the water side, 
where  $\av{n_i} 
= 0.006 v_0^{-1}$ for $z>L/2$ 
and $\av{n_i} =0$ for $z<L/2$. 
We used the model H equations 
and the ion diffusion equations. Here, 
surface disturbances 
grow after  apperance of  an 
elecric double layer in Fig.2.

We also predict droplet  instability 
induced by  an antagonistic 
salt. It is of  interest how  
planar or spherical interfaces 
are deformed and proliferated 
into lamellae in the late stage of 
the instability.

\section{Summary and  remarks}

We  have briefly reviewed 
the recent research on  antagonistic salts 
in water-oil solvents. Note that the selective solvation 
has not been accounted for in the previous  
 electrolyte theories,  though 
it yields   a  variety of 
unexplored effects \cite{Bu}. 
Antagonistic salts should be  one of the  most unique 
   entities in these problems, though 
we have  not yet well understood 
 the puzzling observations by Sadakane {\it et. al} 
 \cite{Sadakane,Sada1,Sada2,SadakanePRL}. 
In particular, it is of interest  how addition of an antagonistic salt 
changes the nature of the solvent criticality 
\cite{Nara,Leys,Sada1}, 
where the periodic fluctuations  appear 
for   $\gamma_{\rm p}>1$ as in    Fig.5. .

A new aspect highlighted in this work 
is that  antagonistic salts undergo 
microphase separation  near  solid surfaces 
 in  mixture solvents, as well as across  water-oil 
interfaces. Intriguing  is  then  
the  behavior of antagonistic salts 
around proteins or  Janus particles  
 having hydrophilic and hydrophobic surface parts.

\noindent{\bf Acknowledgments}\\
\noindent 
This work was supported by KAKENHI (No. 25610122) and
Grants-in-Aid for Japan Society for Promotion of Science (JSPS)
Fellows (Grants No. 241799 and 263111).


\end{document}